\documentclass{PoS}

\newcommand{\VEV}[1]{\left\langle #1\right\rangle}

\title{Semileptonic  $\boldmath{B}$ to $\boldmath{D}$ decays at nonzero recoil with 2+1 flavors of improved staggered quarks}

\ShortTitle{Semileptonic $\boldmath{B}$ to $\boldmath{D}$ decays at nonzero recoil}

\author{Si-Wei Qiu\\
        Physics Department, University of Utah, Salt Lake City, UT 84112, USA\\
        E-mail: \email{siwei.qiu@utah.edu}}
\author{\speaker{Carleton DeTar}%
         \thanks{Presented on behalf of Si-Wei Qiu.}\\
        Physics Department, University of Utah, Salt Lake City, UT 84112, USA\\
        E-mail: \email{detar@physics.utah.edu}}
\author{Daping Du\\
        Department of Physics, University of Illinois, Urbana, IL 61801, USA}
\author{Andreas S.\ Kronfeld\\
        Fermi National Accelerator Laboratory, Batavia, IL 60510, USA}
\author{Jack Laiho\\
        SUPA, Department of Physics and Astronomy, University of Glasgow, Glasgow, Scotland, UK}
\author{Ruth S. Van de Water\\
        Fermi National Accelerator Laboratory, Batavia, IL 60510, USA}
\author{(Fermilab Lattice and MILC Collaborations)}

\abstract{The Fermilab Lattice and MILC collaborations are completing
  a comprehensive program of heavy-light physics on MILC (2+1)-flavor
  asqtad ensembles with lattice spacings as small as 0.045 fm and
  light-to-strange-quark mass ratios as low as 1/20. We use the
  Fermilab interpretation of the clover action for heavy valence
  quarks and the asqtad action for the light valence quarks. The
  central goal of the program is to provide ever more exacting tests
  of the unitarity of the CKM matrix. We present preliminary results
  for one part of the program, namely the analysis of the semileptonic
  decay $B \rightarrow D \ell \nu$ at nonzero recoil.}

\FullConference{The 30th International Symposium on Lattice Field Theory\\
   June 24 - 29, 2012\\
    Cairns, Australia}

\begin{document}

\section{ Introduction\label{sec:In}}

The CKM matrix element $|V_{cb}|$ is a key quantity in Standard-Model
tests.  It normalizes the legs of the unitarity triangle.  The
dominant uncertainty in $|V_{cb}|$ comes from theoretical
determinations of the hadronic form factors for $B \to c\ell\nu + \ldots{}$.
The exclusive processes $B \to D \ell \nu$ and $B \to D^* \ell \nu$
\cite{Bernard:2008dn,Laiho:2005ue,Bailey:2010gb} can be studied in
lattice gauge theory.  Here we report on results for $ B \rightarrow D
\ell \nu$.  Lattice calculations at zero recoil typically have the
smallest errors. However, because of the phase space suppression near
zero recoil in $ B \rightarrow D \ell \nu$, experimental errors are
largest there.  Thus, we aim to work at nonzero recoil where the
combined experimental and theoretical error is minimized.  This work
updates our previous report \cite{Qiu:2011ur} with all ensembles now
included in the analysis.

The differential decay rate $d\Gamma({
  B}\rightarrow {D}\ell\overline{\nu})/dq^2$ is, for $m^2_\ell \ll \min(M^2_B,q^2)$,
proportional to $|f_+|^2$ for $\ell = e,\mu$, where for $q=p_B-p_D$,
\begin{equation}
    \langle D(p_D) | \mathcal{V}^\mu | B(p_B) \rangle = 
        f_+(q^2) \left[ (p_B+p_D)^\mu - \frac{M_B^2-M_D^2}{q^2}q^\mu \right]  +
        f_0(q^2) \frac{M_B^2-M_D^2}{q^2} q^\mu \, .
\end{equation}
Here $\mathcal{V}^\mu = \bar b \gamma^\mu c$  is the $b \to c$ vector current
and $f_+$ and $f_0$ are the vector and scalar form factors, respectively.
The alternative form factors $h_+$ and $h_-$ are convenient:
\begin{equation}
    \frac{\langle D(p_D) | \mathcal{V}^\mu | B(p_B) \rangle }{ \sqrt{M_B M_D}} =
        h_+(w)(v + v^\prime)^\mu  + h_-(w) (v-v^\prime)^\mu \, ,
\end{equation}
where $v = p_B/M_B$ and $v^\prime = p_D/M_D$.  They are related to
$f_+$ and $f_0$ through
\begin{eqnarray}
    f_+(q^2) & = & \frac{1}{2\sqrt{r}} \left[ (1+r) h_+(w) - (1-r) h_-(w) \right]\, ,\nonumber \\
    f_0(q^2) & = & \sqrt{r} \left [ \frac{w+1}{1+r} h_+(w) - \frac{w-1}{1-r} h_-(w) \right ], 
\end{eqnarray}
where $r=M_D/M_B$ and $q^2=M_B^2+M_D^2-2wM_BM_D$ or $w = v\cdot v^\prime$.

\begin{figure}
\vspace*{-5mm}
  \begin{tabular}{cc}
  \begin{minipage}{0.43\textwidth}
  \vspace*{9mm}
  \includegraphics[width=\textwidth]{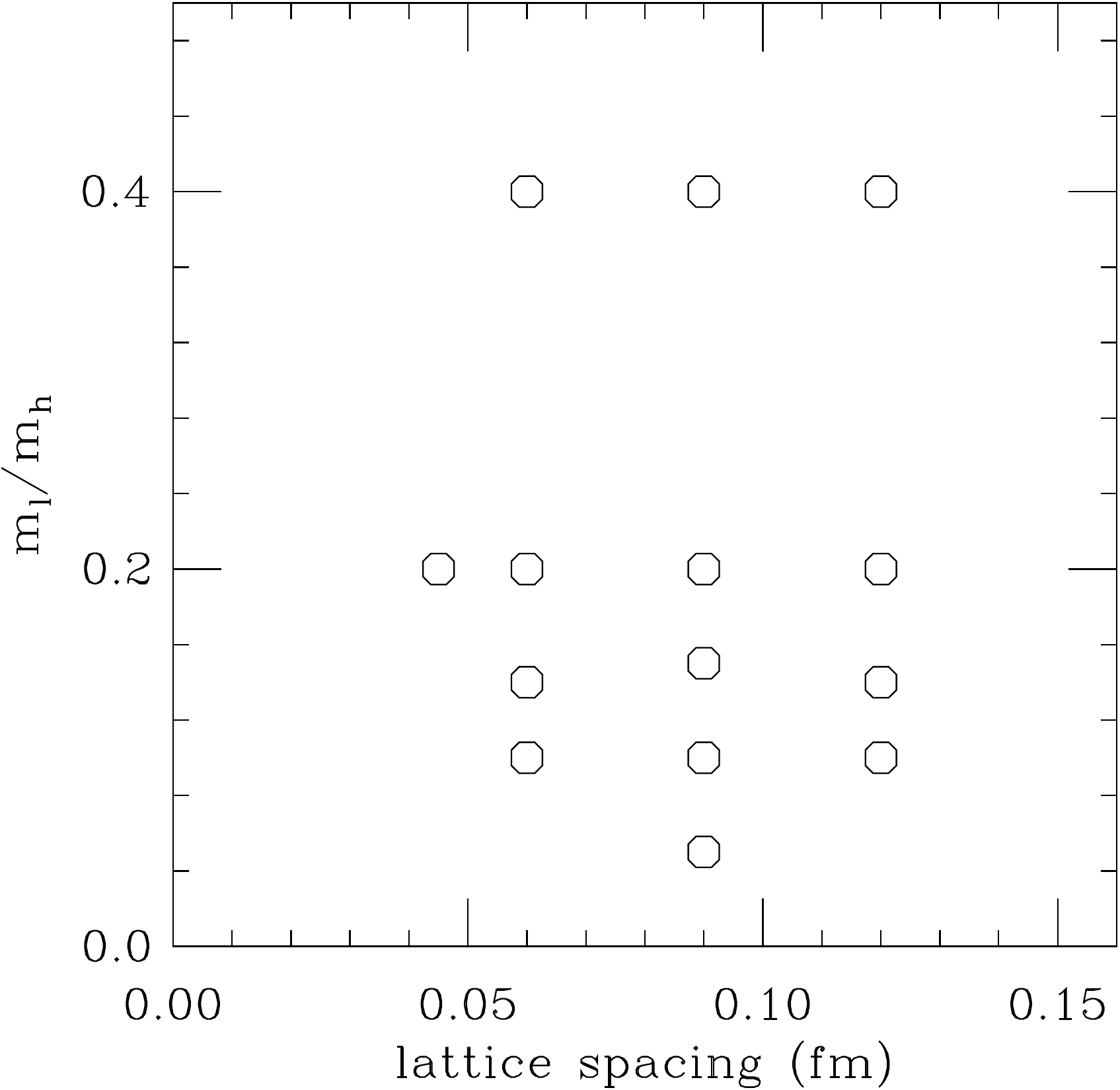}\hfill \\
  \end{minipage}
&
  \begin{minipage}{0.45\textwidth}
  \includegraphics[width=\textwidth]{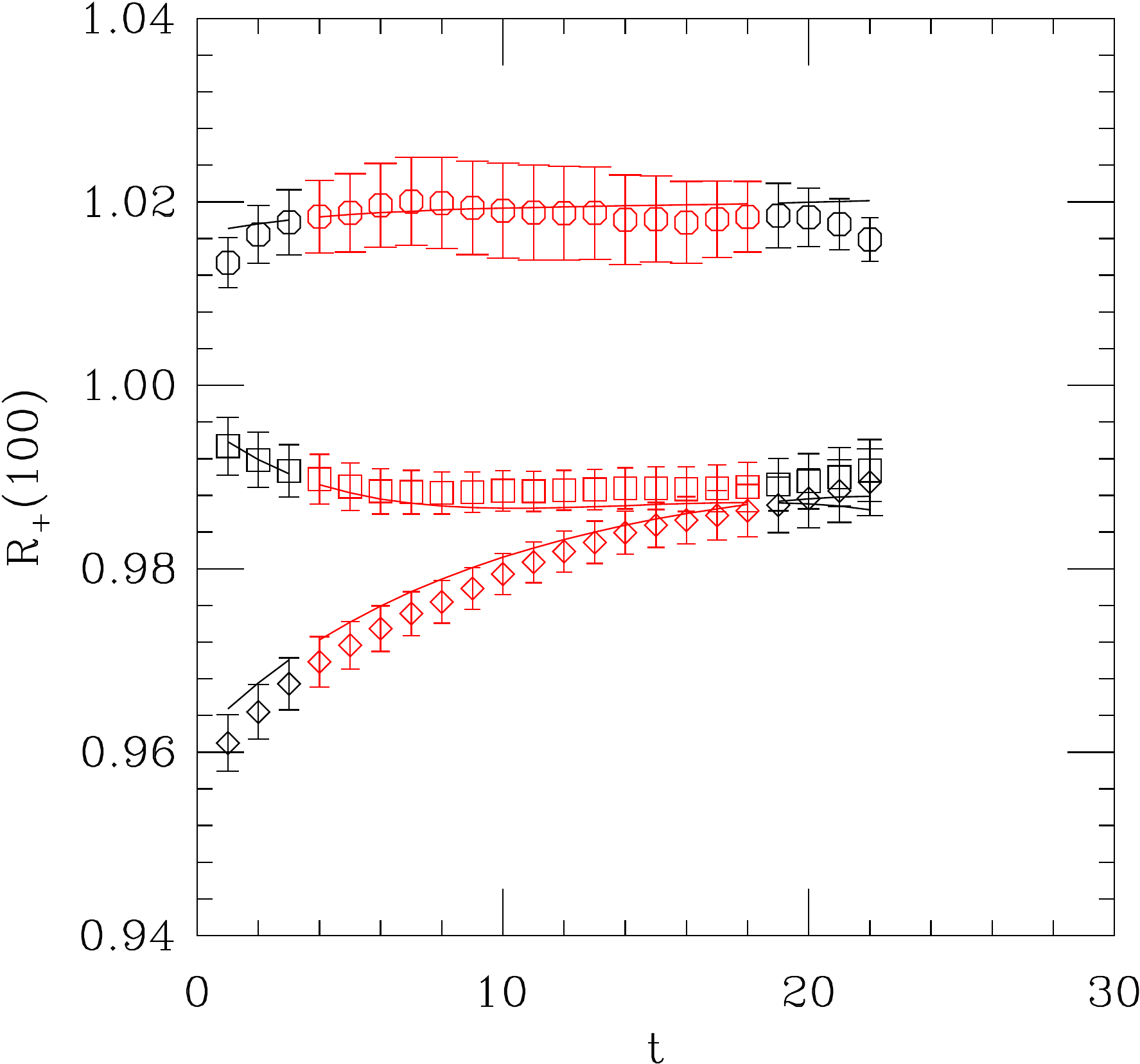}
  \end{minipage}
  \end{tabular}
  \caption{ Left panel: parameters of the 14 lattice ensembles in this
    study. Plotted are values of the light to strange sea quark mass
    ratio $m_\ell/m_h$ {\it vs.} the approximate lattice spacing in
    fm.  Right panel: example from the $a = 0.06$ fm, $m_\ell/m_h =
    0.15$ ensemble with $T =24,25$.  Upper curve:
    the zero-recoil double ratio, middle curve: the ratio $R_+({\bf
      p},t)/R_+({\bf 0},t)$ for the 1S smeared $D$-meson interpolator,
    and bottom curve: the local interpolator. Red points are included
    in the fit ($p = 0.15$).}
\label{fig:R+ensembles}
\end{figure}
\section{Lattice-QCD calculation}

We are carrying out calculations on 14 lattice ensembles generated in
the presence of $2+1$ flavors of improved (asqtad) staggered sea
quarks \cite{ASQTAD} with light-quark masses and lattice spacings
shown in Fig.~\ref{fig:R+ensembles}.

In the $ B$ meson rest frame for any recoil $D$-momentum ${\bf p}$
we can obtain $h_+$ and $h_-$ from matrix elements of the current starting
from ratios of lattice matrix elements $R_+$ and $R_-$, and $x_f$,
where
\begin{eqnarray}
R_+({\bf p})&\equiv&\VEV{D({\bf p})|V^4|B({\bf 0})} \nonumber\\
R_-({\bf p})&\equiv&\frac{\VEV{D({\bf p})|V^1|B({\bf 0})}}{\VEV{D({\bf p})|V^4|B({\bf 0})}} \nonumber \\
x_f({\bf p})&\equiv&\frac{\VEV{D({\bf p})|V^1|D({\bf 0})}}{\VEV{D({\bf p})|V^4|D({\bf 0})}} \\
w({\bf p}) &=& [1+x_f({\bf p})^2]/[1 -x_f({\bf p})^2] \nonumber \\ 
h_+(w)&=&R_+({\bf p})[1-x_f({\bf p})R_-({\bf p})] \nonumber\\
h_-(w)&=&R_+({\bf p})[1-R_-({\bf p})/x_f({\bf p})] \nonumber \, .
\end{eqnarray}
At zero recoil, we can also use the double ratio of Hashimoto {\it et al} \cite{Hashimoto:2001nb}:
\begin{equation}
|h_+({\bf 0})|^2 = \frac{\VEV{D({\bf 0})|V^1|B({\bf 0})}\VEV{B({\bf 0})|V^1|D({\bf 0})}}
  {\VEV{D({\bf 0})|V^4|D({\bf 0})}\VEV{B({\bf 0})|V^4|B({\bf 0})}}\, .
\end{equation}

The continuum $\mathcal{V}^\mu$ and lattice $V^\mu$ currents are
matched through $\mathcal{V}^\mu_{cb} = Z_{V^\mu_{cb}} V^\mu_{cb}$.
We use a mostly nonperturbative method \cite{Hashimoto:2001nb},
writing
\begin{equation}
     Z_{V^\mu_{cb}} = \rho_{V^\mu_{cb}} \sqrt{Z_{V^4_{cc}}Z_{V^4_{bb}}} \, ,
\end{equation}
and determine $\rho_{V^\mu_{cb}}$ from one-loop lattice perturbation theory.

\begin{figure}
  \begin{center}
  \includegraphics[width=0.35\textwidth]{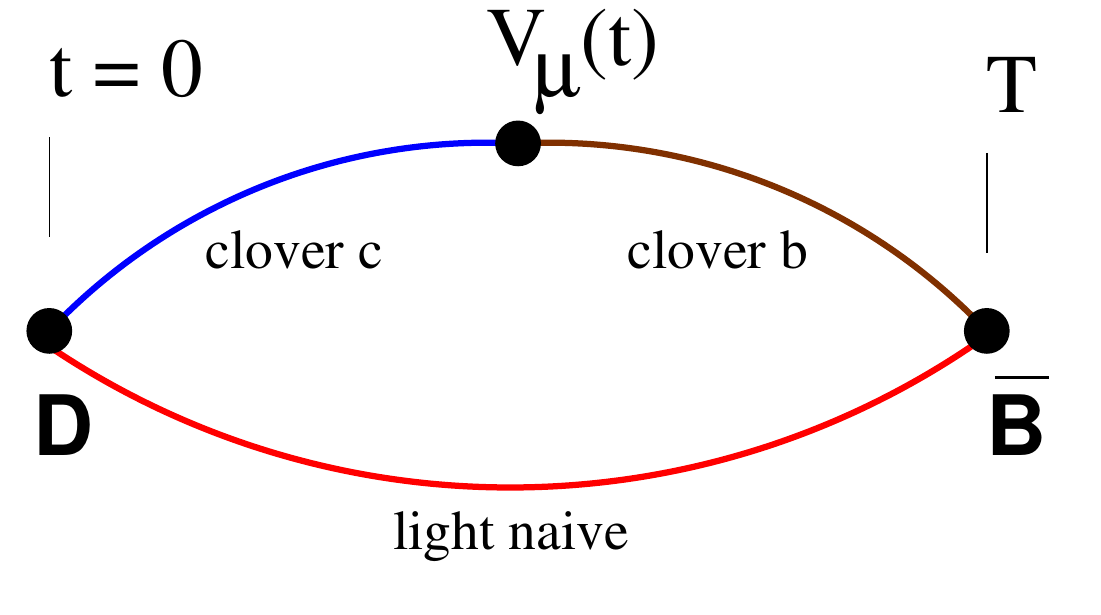}\hfill
  \end{center}
  \vspace*{-5mm}
  \caption{Valence quark line diagram for $B\rightarrow D$.
}
\label{fig:qkline}
\end{figure}

In addition to the two-point functions, we need matrix elements
$\VEV{Y({\bf p})|V^\mu|X(0)}$ for $X,Y \in \{B, D\}$.  They are
constructed from naive light spectator quark propagators and clover
heavy quark propagators in the Fermilab interpretation
\cite{ElKhadra:1996mp} as shown in Fig.~\ref{fig:qkline}. Valence
bottom and charm quark masses were tuned to the ``kinetic'' $B_s$ and
$D_s$ masses.  The mass of the naive light spectator quark is set
equal to that of the light sea quark.  

For interpolating operators ${\cal O}_X$, we compute two-point and
three-point functions
\begin{eqnarray}
   C^{2pt,X}({\bf p},t) &=& \VEV{{\cal O}^\dagger_X(0) {\cal O}_X(t)} \, , \\
   C^{3pt,X\rightarrow Y,\mu}({\bf p};0,t,T) &=& \VEV{{\cal O}_Y^\dagger(0) V^\mu(t) {\cal O}_X(T)} \, .
\end{eqnarray}
We use both point and $1S$ smeared interpolating operators for the $D$ meson and
$1S$ smeared interpolating operators for the $B$.

\subsection{Two-point and three-point correlator fits}

We obtain the lattice form factors via a two-step procedure.  First,
we fit the $B$- and $D$-meson two-point correlators to obtain the
energies and overlap factors.  Then we use these determinations as constraints (with
Bayesian priors) in the three-point fits.  For illustration, we show
the reduction of the three-point and two-point functions to obtain
$R_+(p) = \VEV{D({\bf p})|V^4|B({\bf 0})}$.  We include excited $B$
and $D$ contributions indicated with a prime but not both together:
\begin{eqnarray}
C_{V4}^{3pt,  B\rightarrow D}({\bf p},t)&=&\sqrt{Z_D({\bf p})}\frac{e^{-E_D t}}{\sqrt{2E_D}}
\VEV{D({\bf p})|V^4|B({\bf 0})}\frac{e^{-m_B(T-t)}}{\sqrt{2m_B}}\sqrt{Z_B({\bf 0})} \nonumber \\
&+&\sqrt{Z_{D^\prime}({\bf p})}\frac{e^{-E_{D^\prime}t}}{\sqrt{2E_{D^\prime}}}
\VEV{D^{\prime}({\bf p})|V^4|B({\bf 0})}\frac{e^{-m_B(T-t)}}{\sqrt{2m_B}}\sqrt{Z_B({\bf 0})} \\
&+&\sqrt{Z_D({\bf p})}\frac{e^{-E_Dt}}{\sqrt{2E_D}}
\VEV{D({\bf p})|V^4|B^{\prime}({\bf p})}\frac{e^{-m_{B^\prime}(T-t)}}{\sqrt{2m_{B^\prime}}}\sqrt{Z_{B^{\prime}}({\bf 0})} 
\, ,\nonumber
\end{eqnarray}
or
\begin{equation}
   C_{V^4}^{3pt,  B\rightarrow D}({\bf p},t) =  C_0({\bf p})\VEV{D({\bf p})|V^4|B({\bf 0})}e^{-E_D t}e^{-m_B(T-t)}
    \left[1 + C_1({\bf p})e^{-\Delta E_D t}+C_2({\bf p})e^{(t-T)\Delta m_B}\right] \, ,
\end{equation}
where $C_0({\bf p})$, $\Delta E_D = E_{D^\prime} - E_D$, and $\Delta
m_B = m_{B^\prime} - m_B$ come from fits to two-point
correlators. Terms oscillating as $(-)^t$ (not shown) are 
introduced by the naive light quark.  We suppress their contributions
by averaging over $T$, $T+1$ and $t$, $t+1$, as introduced in
\cite{Bernard:2008dn}.

Putting information from three- and two-point functions together, we get
 \begin{eqnarray}
   R_+({\bf p},t) &\equiv & \frac{C_{V^4}^{3pt,  B\rightarrow D}({\bf p},t)e^{(E_D-m_B)t + (m_B-m_D)T/2}}
         {\sqrt{C_{V^4}^{3pt,D\rightarrow D}({\bf 0},t)C_{V^4}^{3pt,B\rightarrow B}({\bf 0},t)}}
    \sqrt{\frac{Z_D({\bf 0})E_D}{Z_D({\bf p})m_D}}  \nonumber \\
    & \approx & R_+({\bf p}) \left[1 + s_1({\bf p})e^{-\Delta E_D t}+s_2({\bf p})e^{(t-T)\Delta m_B}\right] \, .
 \end{eqnarray}
The zero-recoil form factor $h_+({\bf 0}) = R_+({\bf 0})$ can be
calculated very accurately from the double ratio.  A good strategy is
to use it to normalize the nonzero recoil values:
\begin{equation}
  \frac{R_+({\bf p},t)}{R_+({\bf 0},t)} = \frac{R_+({\bf p})}{R_+({\bf 0})} \exp(\delta m\, t) + 
   A({\bf p}) \exp(-\Delta E_D t) + B({\bf p}) \exp(\Delta m_B t)\, ,
\end{equation}
where $\delta m = 0$, $\Delta E_D =
E_{D^\prime} - E_D$, and $\Delta m_B = m_{B^\prime} - m_B$ are
constrained by fits to two-point functions.

We do a simultaneous fit to three three-point functions, as
illustrated in the right panel of Fig.~\ref{fig:R+ensembles} and
determine $R_+$, $R_-$, and $x_f$ for each momentum (recoil parameter
$w$), from which we determine $h_+$ and $h_-$.

\begin{figure}
\vspace*{-5mm}
  \begin{tabular}{cc}
\hspace*{-10mm}
  \begin{minipage}{0.73\textwidth}
\includegraphics[width=\textwidth]{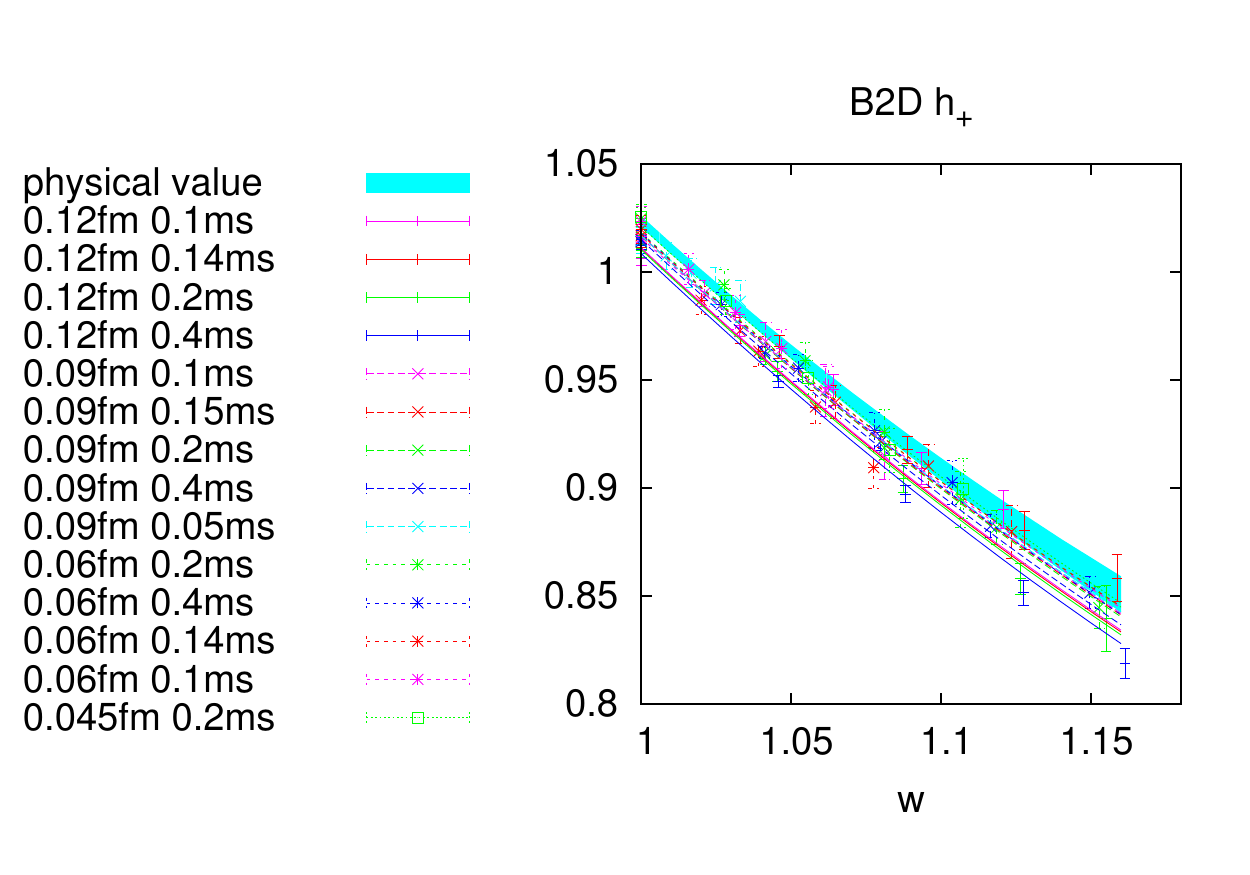}
  \end{minipage}
&
\hspace*{-25mm}
  \begin{minipage}{0.63\textwidth}
\includegraphics[width=\textwidth]{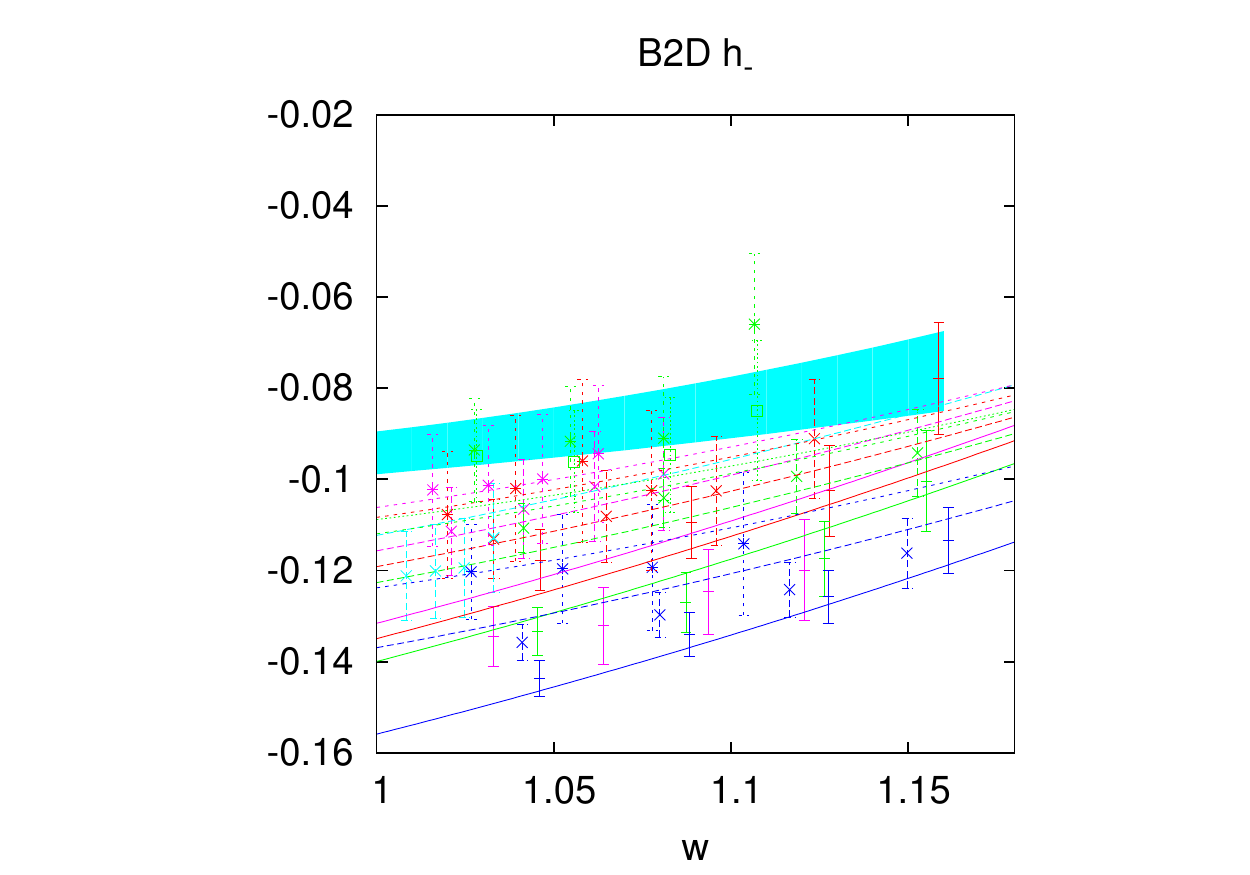}
  \end{minipage}
  \end{tabular}
  \caption{Form factors $h_+$ (left) and $h_-$ (right)
    as a function of the recoil parameter $w$. The curves in
    each panel are a result of a simultaneous fit to the
    chiral-continuum expressions of Eq.~(\protect\ref{eq:chiral}) with
    $p = 0.68$ for the $h_+$ fit and $p = 0.41$ for the $h_-$ fit.}
\label{fig:h+h-}
\end{figure}

\subsection{Chiral-continuum extrapolation and $q^2$ parameterization}

The resulting form factors $h_+$ and $h_-$ are shown in
Fig.~\ref{fig:h+h-}.  We fit them to the expressions
\begin{eqnarray}
  h_+(a,m_\ell,w) & = & 1 - \rho_+^2 (w-1) + k_+ (w-1)^2 + \frac{ X_+(\Lambda_\chi)}{m_c^2}
                 + c_{1,+} m_\ell + c_{a,+} a^2 + c_{a,w,+} a^2(w-1) \nonumber \\
                 & + & \frac{g_{D^*D\pi}^2}{16 \pi^2 f^2} {\rm logs}_{\rm 1-loop}(\Lambda_\chi,w,m_\ell,a) 
    \label{eq:chiral}\\
  h_-(a,m_\ell,w) &=& \frac{X_-}{m_c} - \rho_-^2 (w-1) + k_-(w-1)^2 
       + c_{1,-} m_\ell + c_{a,-} a^2  +  c_{a,w,-} a^2(w-1) \nonumber
\end{eqnarray}
for light spectator quark mass $m_\ell$, lattice spacing $a$, and $w =
v\cdot v^\prime$.  For the one-loop chiral logs we use a staggered
fermion version of Chow and Wise \cite{Chow:1993hr}.  Thus, these fit
functions contain the correct next-to-leading-order chiral
perturbation theory expressions, including staggered discretization
effects \cite{Bailey:2012rr}.  As can be seen, the dependence of $h_+$
on $a$ and $m_\ell/m_h$ is quite mild.  We expect $h_-$ to have larger
discretization effects than $h_+$ because of different HQET power
counting.  This is consistent with what we see in the data.  For
$|V_{cb}|$, the contribution coming from $h_-$ over the entire
kinematic range is small, so the larger errors in $h_-$ don't increase
the overall error much.  These features with 14 ensembles are
consistent with our previous findings with four ensembles
\cite{Bailey:2012rr}.

\begin{figure}
\vspace*{-5mm}
  \begin{tabular}{cc}
\hspace*{-25mm}
  \begin{minipage}{0.60\textwidth}
  \vspace*{-6mm}
  \includegraphics[width=\textwidth]{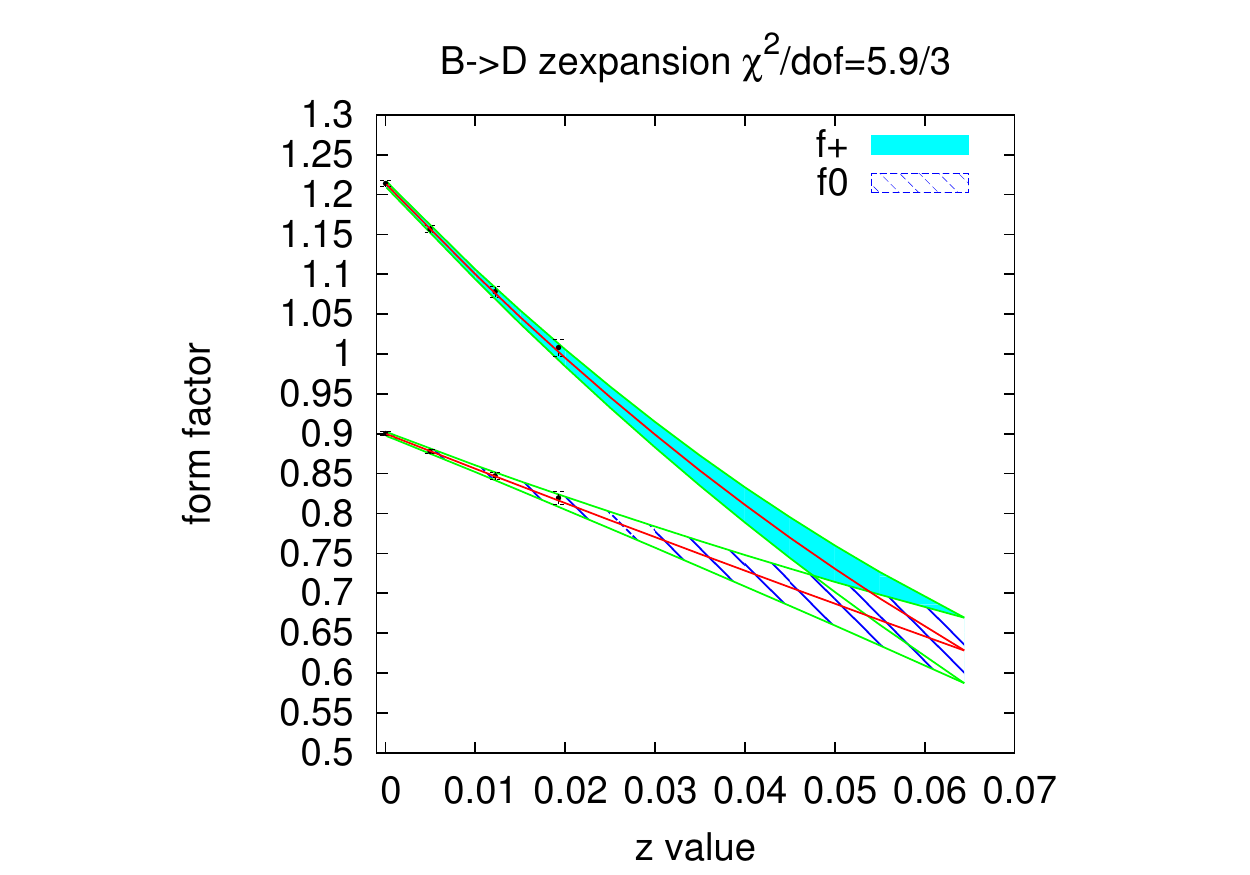}\hfill
  \end{minipage}
&
\hspace*{-15mm}
  \begin{minipage}{0.65\textwidth}
  \vspace*{6mm}
  \includegraphics[width=\textwidth]{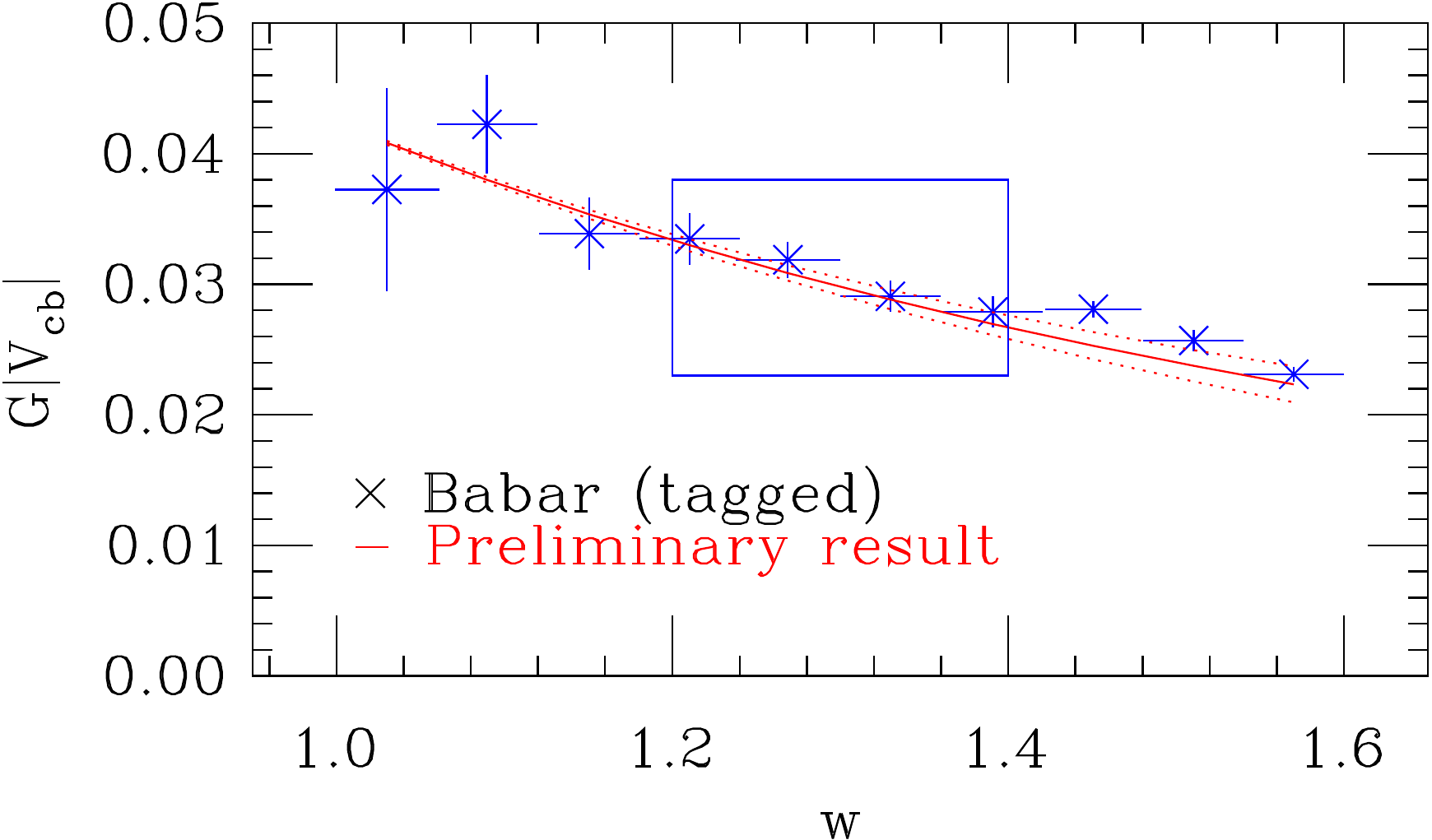}\hfill \\
  \end{minipage}
  \end{tabular}
  \caption{Left: form factors $f_+$ and $f_0$ parameterized by the $z$
    expansion ($p = 0.12$).  Right: comparison with experimental
    results from the Babar collaboration \cite{Aubert:2009ac}.  The
    red line gives our result.  The red dotted lines show only
    statistical errors.  The boxed region appears to have the smallest
    combined error.}
\label{fig:z-comparison}
\end{figure}

To compare the lattice and experimental form factors we need to
extrapolate to larger $w$ (equivalently $q^2$).  We do this using the
$z$-expansion of Boyd, Grinstein and Lebed \cite{Boyd:1994tt}, which
provides a model-independent parameterization of the $q^2$ dependence
of $f_+$ and $f_0$.  This expansion builds in constraints from
analyticity and unitarity.  It is based on the
conformal map
\begin{equation}
   z(w) = \frac{\sqrt{1+w}-\sqrt{2}}{\sqrt{1+w}+\sqrt{2}} \, ,
\end{equation}
which maps the physical region $w \in [1, 1.59]$ to $z \in [0,
  0.0644]$.  It pushes poles and branch cuts far away at $|z| \approx
1$.  Form factors are then parameterized as
\begin{equation}
  f_i(z) = \frac{1}{P_i(z) \phi_i(z)} \sum_{n=0}^{\infty} a_{i,n} z^n \, ,
\end{equation}
where $P_i(z)$ are the Blaschke factors and $\phi_i$ are the ``outer
functions''.  The latter are chosen to simplify the unitarity bound:
\begin{equation}
  \sum_n |a_{i,n}|^2 \le 1 \, .
\end{equation}
In practice, we need only the first few coefficients in the expansion.
We also impose the kinematic constraint $f_+ = f_0$ at $q^2 = 0$ or $z
\approx 0.0644$.

To implement the $z$ expansion, we start from the value of $f_+$ and
$f_0$ at the physical point, as determined from the chiral/continuum fit.
We choose four $w$ values, $w=1.00$, 1.04, 1.10, and 1.16, and use the
corresponding form factor values to determine the coefficients $a_{i,0}$,
$a_{i,1}$, and $a_{i,2}$.  These, then, are used to parameterize the form
factors over the full kinematic range, as shown in
the left panel of Fig.~\ref{fig:z-comparison}.

We compare our result with experimental measurements from the Babar
collaboration \cite{Aubert:2009ac} in Fig.~\ref{fig:z-comparison}. For
present purposes we take $|V_{cb}|$ from $ B \to D^*\ell\nu$ at zero
recoil \cite{Bailey:2010gb}.

\section{\label{sec:Fut} Future plans}

To complete the analysis, we need to apply small corrections resulting
from adjusting the charm and bottom quark masses to their tuned
values, implement the full current renomalization, and compile a
complete error budget.

\section*{Acknowledgements} Computations for this work were carried
out with resources provided by the USQCD Collaboration, the National
Energy Research Scientific Computing Center and the Argonne Leadership
Computing Facility, which is funded by the Office of Science of the
U.S. Department of Energy; and with resources provided by the National
Institute for Computational Science and the Texas Advanced Computing
Center, which are funded through the National Science Foundation's
Teragrid/XSEDE Program. This work was supported in part by the
U.S.\ Department of Energy under grant No.\ DE-FG02-91ER40677 (D.D.)
and the U.S.\ National Science Foundation under grants PHY0757333 and
PHY1067881 (C.D.) and PHY0903571 (S.-W.Q.). J.L.\ is supported by the
STFC and by the Scottish Universities Physics Alliance. This
manuscript has been co-authored by employees of Brookhaven Science
Associages, LLC, under Contract No. DE-AC02-98CH10886 with the
U.S.\ Department of Energy. Fermilab is operated by Fermi Research
Alliance, LLC, under Contract No.\ DE-AC02-07CH11359 with the United
States Department of Energy.

\providecommand{\href}[2]{#2}

\end{document}